\documentclass[a4paper,fleqn,usenatbib]{mnras}

\usepackage{newtxtext,newtxmath}

\usepackage[T1]{fontenc}
\usepackage{ae,aecompl,multirow}
\usepackage{hyperref}

\makeatletter
\def\footnoterule{\kern-3\p@
  \hrule \@width 2in \kern 2.6\p@} 
\makeatother

\usepackage{graphicx}	
\usepackage{amsmath}	
\usepackage{amssymb}	
\usepackage{subfig}

\usepackage{caption}
\title[X-ray spectral curvature of MKN\,421]{Influence of Energy-Dependent Particle Diffusion on the X-ray spectral
curvature of MKN\,421}

\author[P. Goswami et al.]{Pranjupriya Goswami$^{1}$\thanks{E-mail: pranjupriya.g@gmail.com},
Sunder Sahayanathan$^{2}$,
Atreyee Sinha$^{3,4}$, 
Ranjeev Misra$^{3}$ \newauthor
and Rupjyoti Gogoi$^{1}$
\\
$^{1}$Department of Physics, Tezpur University, Napaam - 784028, India\\
$^{2}$Astrophysical Sciences Division, Bhabha Atomic Research Centre, Mumbai - 400085, India\\
$^{3}$Inter-University Center for Astronomy and Astrophysics, Post Bag 4, Ganeshkhind, Pune - 411007, India\\
$^{4}$APC, AstroParticule et Cosmologie, Universit\'e Paris Diderot, CNRS/IN2P3, CEA/Irfu, Observatoire de Paris,\\
Sorbonne Paris Cit\'e, 10, rue Alice Domon et L\'eonie Duquet, 75205 Paris Cedex 13, France
}

\date{Accepted XXX. Received YYY; in original form ZZZ}

\pubyear{2018}

\begin{document}
\label{firstpage}
\pagerange{\pageref{firstpage}--\pageref{lastpage}}
\maketitle

\begin{abstract}
The X-ray spectral curvature of blazars is traditionally explained by an
empirical log-parabola function characterized by three parameters, namely the flux,
curvature and spectral index at a given energy. Since 
their exact relationship with the underlying physical quantities is unclear, 
interpreting the physical scenario of the source through these parameters is difficult. 
To attain an insight on the X-ray spectral shape, we perform a detailed 
study of the X-ray spectra of the blazar MKN\,421, using an analytical model
where the electron diffusion from the particle acceleration site is energy-dependent. 
The resultant synchrotron spectrum is again determined by three parameters, 
namely, the energy index of the escape time scale, the quantity connecting 
the electron energy to the observed photon energy and the normalization. The 
X-ray observations of MKN\,421, during July 2012 -- April 2013 by \emph{Nu}STAR 
and \emph{Swift}-XRT are investigated using this model and we find a significant 
correlation between model parameters and the observational quantities. 
Additionally, a strong anti-correlation is found between the fit parameters defining the
spectral shape, which was not evident from earlier studies using empirical 
models. This indicates the flux variations in MKN 421 and possibly other blazars, may arise from a definite physical process that needs to be further
investigated.

\end{abstract}



\section{Introduction}
Blazars are an extreme class of Active Galactic Nuclei (AGN) with the relativistic jet aligned towards the observer. 
Their broadband spectral energy distribution (SED), extending from radio to $\gamma$-ray energies, is characterized by two 
prominent peaks with the low energy component attributed to synchrotron emission and the high energy to inverse Compton
processes. 
The spectral modelling of the synchrotron component suggests the underlying electron distribution to be a broken power-law with a smooth 
transition at the break energy corresponding to the peak of the SED \citep{sambruna, Tavecchio1998, Fossati2000}. Alternatively, the spectrum around the peak is 
well reproduced by a log-parabola function, requiring a lesser number of free parameters 
\citep{Landau1986, Massaro2004a, Tramacere2007, Paggi2009, chen2014curvature}. 
Recent high-resolution optical -- X-ray observations showed that the spectrum  deviates from a power-law even beyond 
the synchrotron peak and is better represented by 
functions with mild curvature \citep{bhagwanPKS2014,gaurPKS2017}.
These observational inputs motivate one to look for a modified non-thermal electron distribution
capable of explaining the broadband synchrotron spectra of blazars.

The log-parabola spectral shape of the synchrotron peak was observed to be the best-fit model for various blazars and during
different flux states. This initiated an extensive study of the peak spectral curvature to identify 
possible correlations between  the model parameters and to relate their property to the underlying particle 
acceleration process \citep{Massaro2004a, Massaro2004b, Massaro2008, Tramacere2007}. For most of 
the sources, the curvature parameter is positively correlated with the flux corresponding to peak energy, while
no substantial correlation was observed between the curvature parameter and 
the spectral index at a fixed energy \citep{kapanadzePKS2014, atreyeeMkn421}. On the other hand, several studies show a significant
positive correlation between the peak energy of the fitted log-parabola function and the corresponding flux \citep{Massaro2006, Massaro2008, Tramacere2007, Paggi2009}.
The dependence of the peak energy to the corresponding flux can be used to identify the dominant parameter responsible for the 
spectral change. However, these studies are often performed over a narrow  X-ray energy band where a log-parabola function provides a decent fit statistic.

A log-parabola photon spectral shape can originate if the emitting electron distribution is also a log-parabola. 
Such a distribution can originate from an energy dependence of the acceleration probability, as shown by \citep{Massaro2004a}. 
Alternatively, the curvature can also be 
interpreted as a result of the energy dependence of the escape time scale from the emission region. This model is capable of 
explaining the broadband SED with a power-law spectral shape at low energies and a smooth curvature at the high energies \citep{atreyeeCurvature1ES}. 
On the other hand, an energy dependence of the escape time scale in the acceleration region is capable of producing a spectrum that
deviates from a power-law and closely agrees with a broken log-parabola \citep{Sitha2018}.

The blazar MKN\,421 is one of the nearest (z=0.031) and the brightest known  BL Lac object that has been studied extensively with multi-epoch observations over
 a wide spectral range \citep{2009ApJ...703..169A, Abdo2011421, 2012A&A...542A.100A, asinhahagar, 2016ApJS..222....6B}. This source is classified as a high 
frequency BL Lac (HBL) as its synchrotron peak falls at X-ray energies \citep{Padovani1995}. The X-ray spectra of this source exhibit significant curvature in both flaring and quiescent states. The high flux state spectra can be well fitted by a log-parabola function \citep{Massaro2004a, Massaro2006, Donnarumma2009, dahai2013, atreyeeMkn421}; whereas, the low flux state spectra are better explained with power-law or power-law with exponential cut-off models \citep{dahai2013, kataoka2016}. 
Alternatively,  \cite{Qzhu2016} showed that the X-ray and $\gamma$-ray spectra of the source can be reproduced using synchrotron and self-Compton emission models
with the underlying electron distribution assumed to be a combination of power-law and log-parabola shape.
A detailed X-ray spectral study of the source, at energies 0.1 -- 100 keV, was
 performed by \citet{Massaro2004a} using BeppoSAX observations of May 1999 and April -- May 2000. The X-ray spectra can be well fitted by a log-parabola
 function with the curvature parameter showing a strong positive correlation with the spectral index at 1 keV. However, no such correlation between these
 model parameters is found in a combined \emph{Swift}-XRT and \emph{Nu}STAR spectra of the source during April 2013 \citep{atreyeeMkn421}. Similar to other
 BL Lac objects, MKN\,421 also showed a strong positive correlation between the log-parabola peak energy and the peak flux and a negative correlation between
 peak energy and the curvature parameter \citep{Massaro2004b, Massaro2006, Massaro2008, Tramacere2007, Tramacere2009}. Interestingly, the flux to hardness
 ratio plot of the source during various flaring episodes shows both clockwise or anticlockwise patterns \citep{Tramacere2009, Kapanadze4212016}. 

In the present work, we analyse \emph{Nu}STAR observations of the blazar MKN 421 during the period July, 2012 to April, 2013 and perform a detailed 
investigation of the X-ray spectral curvature. Additionally, we supplement this with \emph{Swift}-XRT observation
at low energies, thereby enabling a broadband (0.3-80 keV) X-ray study of the source. 
To gather a better insight into the physics behind the X-ray spectral curvatures, we use a physically motivated model by considering the 
dynamics of the electrons in the main acceleration zone. The broadband X-ray spectrum is fitted using this analytical model and the
results are presented. Finally, we study the correlations between different model parameters and compare them with the 
ones obtained from the empirical log-parabola spectral fit. 

The paper is organised as follows: In the following section, we describe the details of
the data analysis procedure and in Section \ref{analysis}, we present our spectral study using the empirical log-parabola model. The details 
about the analytical model involving the dynamics of the particle acceleration and the spectral fitting are presented in Section \ref{engtescmodel} along with
the correlation studies. Finally, in Section \ref{discussion} we discuss our results in light of current scenario of the blazar spectral studies.

\section{X-ray observation and analysis}
\label{observations}

MKN\,421 has been monitored by \emph{Nu}STAR in the hard X-ray energy band during  
flaring as well as quiescent flux states. The source underwent a strong X-ray flare from 10 -- 19 April 2013 and was observed by both \emph{Nu}STAR 
and \emph{Swift}-XRT simultaneously. In order to study the spectral behaviour of the source 
during different flux states, we selected 20 \emph{Nu}STAR pointing of MKN\,421, spread over both flaring and non-flaring states. For 10 of these selected epochs, simultaneous
\emph{Swift}-XRT observations were available and they are included in the present study. This enables us to investigate the source over a broad X-ray energy
ranging from 0.3 -- 79 keV. The details of these observations are given in Table \ref{observation} and the data reduction techniques 
used are described below.

\begin{table*}
\centering
\footnotesize
\caption{Details of \emph{Nu}STAR observations for 20 different pointings and the simultaneous \emph{Swift}-XRT observations}
\label{observation}
\vspace{0.3cm} 
\begin{tabular}{lcccccccr}
\hline \hline
Obs.&\multicolumn{3}{c}{\emph{Nu}STAR } &
\multicolumn{3}{c}{\emph{Swift}-XRT} & \\
\cline{2-4}
\cline{6-9}

&Obs. ID& Obs. date $\&$ time &Exposure &&Obs. ID & Obs. date $\&$ time&Exposure \\
&       & (dd-mm-yy)          & (sec)   &&        & (dd-mm-yy)         & (sec) \\
\cline{2-4}
\cline{6-9}

s1 &10002015001 & 2012-07-07 T01:56:07& 	42034&& -- & -- &  --& \\
s2 &10002016001 &	2012-07-08 T01:46:07&24885& &-- & -- & 	--&\\
s3 &60002023006 &2013-01-15 T00:56:07& 24182& &-- & -- &  	--& \\
s4 &60002023010 &2013-02-06 T00:16:07& 19307& &-- & -- &  	--&\\
s5 &60002023012 &	2013-02-12 T00:16:07& 	14780& &-- & -- &  	--&\\
s6 &60002023014 &2013-02-16 T23:36:07& 17359& &00080050006 &2013-02-17 T00:03:59	&9201 	&\\
s7 &60002023016 &2013-03-04 T23:06:07& 17252&& 00080050007 & 2013-03-04 T23:34:25	&	984&\\
s8 &60002023018 &2013-03-11 T23:01:07& 17474&& 00080050011 & 2013-03-11 T23:58:59& 8425	&\\
s9 &60002023020 &2013-03-17 T00:11:07&16558 && -- & -- &  --&\\
s10&60002023022 &2013-04-02 T17:16:07&24772 & &-- & -- &  	--&\\
s11&60002023024 &2013-04-10 T21:26:07&5758& &-- & -- & --& \\
s12&60002023025 &2013-04-11 T01:01:07& 	57509&& 00080050016 & 2013-04-11 T00:30:59&1076	&\\
s13&60002023026 &2013-04-12 T20:11:07& 	441& &00080050019 & 2013-04-12 T21:53:58&9546&\\
s14&60002023027 &2013-04-12 T20:36:07& 7630& & 00080050019& 2013-04-12 T21:53:58&9546&\\
s15&60002023029 &2013-04-13 T21:36:07& 	16510& &-- & -- & 	--&\\
s16&60002023031 &	2013-04-14 T21:41:07& 15606& &-- & --  & 	--&\\
s17&60002023033 &2013-04-15 T22:01:07& 17278& &00035014062  & 2013-04-15 T23:07:59&534&\\
s18&60002023035 &2013-04-16 T22:21:07& 20279& &00035014065 & 2013-04-17 T00:46:59&8842&\\
s19&60002023037 &2013-04-18 T00:16:07& 17795& &00035014066 & 	2013-04-18 T00:49:59& 6887	&\\
s20&60002023039 &2013-04-19 T00:31:07&15958 & &00035014067  & 	2013-04-19 T00:52:59 & 	6132&\\
\hline \hline  
\end{tabular} \\
\textbf{Notes:} \emph{Nu}STAR observations of 20 different flux states during July 2012 to April 2013 and near simultaneous \emph{Swift}-XRT observations  \\
\end{table*}

\smallskip

\subsection{\emph{Nu}STAR}
\label{nu}

Nuclear Spectroscopic Telescope Array (\emph{Nu}STAR) is a space-based focusing high energy 
X-ray telescope, consisting of two co-aligned, independent telescopes; FPMA and FPMB (Focal Plane Module A and B) 
for parallel observations \citep{Harrison2013}. It operates in the hard X-ray energy band from 3 to 79 keV with subarcmin angular 
resolution and  provides a more than 100-fold improvement in sensitivity beyond 10 keV over other telescopes. \emph{Nu}STAR archival data 
were obtained from NASA's HEASARC interface\footnote{\href{https://heasarc.gsfc.nasa.gov/}{https://heasarc.gsfc.nasa.gov/}} and the data were reduced 
using the \emph{NuSTARDAS} software  package (Version 1.4.1) built in \emph{HEAsoft} (Version 6.19). A circular region of radius $30"$ and centred around 
the source is used to extract the source spectrum and the background is estimated from a circular region of radius $50"$
free from source contamination but adjacent to it. 
With these selections of source and background regions, the \emph{nuproduct} (Version 0.2.8) script was 
used to produce the spectrum for each observation \footnote{\href{https://heasarc.gsfc.nasa.gov/docs/nustar/analysis/}
{https://heasarc.gsfc.nasa.gov/docs/nustar/analysis/}}. 
Finally, the clean event files, ancillary response files (ARFs) and response 
matrices (RMF) were  generated using the standard \emph{nupipeline} (Version 0.4.5). 
The source spectrum for FPMA and FPMB were then grouped individually using the  \emph{grppha} tool to have a minimum of 30 counts/bin. This ensures $\chi^{2}$ statistics will remain valid during spectral fitting using XSPEC \citep{Arnaud96}.

\subsection{\emph{Swift}-XRT}
\label{sw}

The \emph{Swift}-XRT \citep{burrowsSwift} observations are obtained from NASA's HEASARC interface and the data reduced using the \emph{XRTDAS} software package (Version 3.0.0) built-in \emph{HEAsoft}. The observations were made 
in Windowed Timing (WT) mode and standard procedures followed to calibrate and generate the cleaned event files using \emph{xrtpipeline} (Versionv.0.13.0) \footnote{\href{http://www.swift.ac.uk/analysis/xrt/index.php}{http://www.swift.ac.uk/analysis/xrt/index.php}}. To extract the source spectrum, a circular region of 20 pixels centred at the source
is used and for the background, a nearby 40 pixels circular region is selected which is free from the source contamination.
In case of observations with pile up, an annular region with 2 pixels inner boundary and 20 pixels outer boundary is used. 
The spectrum is produced using \emph{xrtproducts} (Versionv 0.4.2) and the ancillary response files (ARFs) files and the 
response matrices (RMF) were generated using  \emph{xrtmkarf} from the \emph{Swift} CALDB. The 0.3 -- 10 keV spectra were then 
grouped as described in the previous section. 

\subsection{X-ray Spectral Analysis}
\label{analysis}

We repeat the analysis of \cite{atreyeeMkn421} by fitting the X-ray spectra due to \emph{Nu}STAR and \emph{Swift}-XRT observations 
of MKN\,421 using a log-parabola model, in-built in XSPEC. This will let us to compare the effectiveness of the analytical model
given in \S \ref{engtescmodel} with the empirical one presented here.
A log-parabola model is defined as \citep{Massaro2004a} 
\begin{align}\label{eq:logpar}
F(E)=K \left(\frac{E}{E_*}\right)^{-\alpha-\beta\,\rm{log}(E/E_*)}
\end{align}
where, $\alpha$ is the spectral index at energy $E_*$, $\beta$ is the curvature parameter and $K$ is the normalization.
Choosing $E_*=1$ keV, the final spectrum is defined by three parameters $\alpha$, $\beta$ and $K$. The knowledge of 
the parameters $\alpha$ and $\beta$ lets us estimate the peak of the log-parabola function in $E^2 F(E)$ representation as
\begin{align}\label{eq:lppeak}
	\rm{log}\,\left(\frac{E_{p,lp}}{E_*}\right)=\frac{2-\alpha}{2\beta}
\end{align}

The hard X-ray spectra of MKN\,421, in the energy range 3 -- 79 keV corresponding to 20 observations by \emph{Nu}STAR, 
is fitted using a log-parabola model and the results are presented in Table \ref{nustar} (left). The Galactic neutral 
hydrogen absorption in the direction of MKN\,421 is estimated as 1.92 $\times$ 10$^{20}$ cm$^{-2}$ \citep{Kalberla2005} 
and fixed throughout the analysis for all the observations. Similar to the earlier results, we found this model can represent the
hard X-ray spectra of all the epochs well with a significant curvature ($\beta$ ranging from $0.14$ to $0.49$). The 
hard X-ray spectra mainly reflects the falling part of the synchrotron spectral component and misses the curvature close to the 
peak. To verify whether the same model is capable of explaining this spectral regime as well, we repeat the spectral fitting
by including simultaneous soft X-ray spectra available from \emph{Swift}-XRT observations for 10 epochs (Table \ref{observation}).
We find that the quality of the fits slightly deteriorate, with a significant change in the spectral parameters and
in Table \ref{swnu} (left), we provide the fit results of this combined \emph{Swift}-XRT--\emph{Nu}STAR observations. 
Interestingly, it can also be noted that the spectral fit is better for the low flux states compared to the high flux states. 
In Table \ref{e_p} (left), we provide the SED peaks obtained using the log-parabola fit parameters corresponding to  \emph{Nu}STAR
alone and \emph{Swift}-XRT -- \emph{Nu}STAR combined spectral fits. 

\begin{figure*}
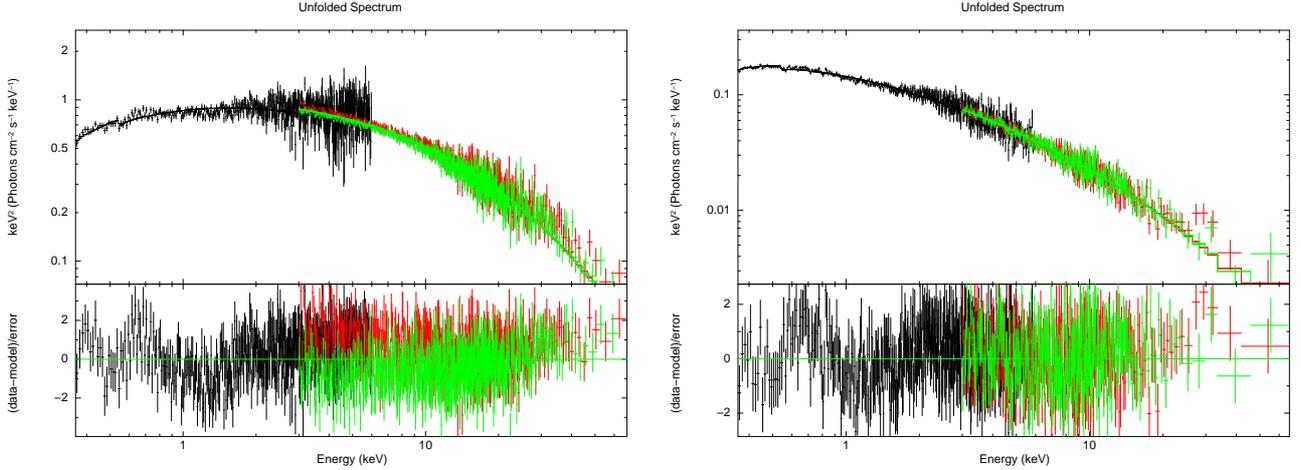

\centering

	\includegraphics[scale=0.34, angle=270]{19_27_engtesc}
	\includegraphics[scale=0.34, angle=270]{11-18_sim}
	
	\caption{Spectral fit with simultaneous \emph{Swift}-XRT (black) and \emph{Nu}STAR (FPMA: red and FPMB: green) observations corresponds to high flux s14(LHS) and low flux s8(RHS) state observations using our energy-dependent electron diffusion model.}
	\label{engtescplot}
\end{figure*}

\section{Energy-dependent Electron Diffusion Model}
\label{engtescmodel}
A better understanding of the MKN\,421 spectrum and its evolution can be perceived by modelling the X-ray spectra using
a physically motivated model rather than a simple empirical function given by equation (\ref{eq:logpar}).
Accordingly, we consider the electron acceleration and the emission processes in the 
blazar jet by dividing the 
main emission zone into two sub regions namely, acceleration region (AR) and cooling region (CR). AR is assumed to be
located around a shock front where electrons are accelerated to relativistic energies and diffuse into 
CR where they lose their energy through radiative processes. Considering a mono-energetic electron injection ($Q_o$) into the AR, the electron distribution in AR ($n_{ar}$) and CR ($n_{cr}$) after attaining steady state can be described as \citep{Kardashev1962},

 
\begin{align}\label{eq:AR}
	{\rm AR:}&\quad\frac{d}{d\gamma} \left[\left(\frac{\gamma}{\tau_a}-A\gamma^2\right)n_{ar}(\gamma)\right] + \frac{n_{ar}(\gamma)}{\tau_e} = Q_o\delta(\gamma-\gamma_0) \\
	\label{eq:CR}
	{\rm CR:}&\quad-\frac{d}{d\gamma} \left[B\gamma^2 n_{cr}(\gamma)\right] + \frac{n_{cr}(\gamma)}{t_e} = \frac{n_{ar}(\gamma)}{\tau_e}
\end{align}
where, $\gamma$ is the dimensionless electron energy ($E=\gamma m_e c^2$), $\tau_a$ and $\tau_e$ is the characteristic 
electron acceleration and escape timescales from the AR, $t_e$ is the escape time scale from CR
and, $A\gamma^2$ and $B\gamma^2$ determine the radiative energy loss rates in AR and CR happening in the 
Thomson regime of the scattering process. 
If we assume a constant $\tau_a$ and an energy-dependent $\tau_e$ such that their ratio takes the form
\begin{align}
	\frac{\tau_a}{\tau_e} = \eta_0\gamma^\kappa
\end{align}
then the electron distribution in AR will be \citep{Sitha2018}
\begin{align}\label{eq:arsoln}
	n_{ar}(\gamma)\propto \gamma^{-1} \rm{exp}\left(-\frac{\eta_0}{\kappa}\,\gamma^{\kappa}\right)
\end{align}
Substituting equation (\ref{eq:arsoln}) in (\ref{eq:CR}), we can obtain the cooled electron distribution in CR as
\begin{align}
	n_{cr}(\gamma)\propto \gamma^{-2}\rm{exp}\left(-\frac{\eta_0}{\kappa}\,\gamma^{\kappa}\right)\quad\rm{for}\quad \gamma\gg\frac{1}{Bt_e}
\end{align}

The synchrotron emissivity at photon energy $E'$ due to the electron distribution in CR can be obtained as 
\footnote{$E'$ is the photon energy measured in CR frame}
\begin{align}\label{eq:syn_emiss}
	j_{\rm syn}(E')= \frac{1}{4\pi}\int P_{\rm syn}(\gamma,E')\,n_{cr}(\gamma)\,d\gamma
\end{align}
where, $P_{\rm syn}$ is the single particle synchrotron emissivity \citep{Rybicki_Lightman}.
Consequently, the flux at the observer can be estimated considering the relativistic Doppler boosting 
from the blazar jet and cosmological effects as \citep{1984RvMP...56..255B,1995ApJ...446L..63D}
\begin{align}\label{eq:obs_flux}
	F_{\rm syn}(E)= \frac{\delta_D^3(1+z)}{d_L^2} V 
	j_{\rm syn}\left(\frac{1+z}{\delta_D}E\right)
\end{align}
Here, $\delta_D$ is the jet Doppler factor, $z$ is the redshift of the source, $d_L$ is the luminosity distance 
and $V$ is the volume of CR. Approximating $P_{\rm syn}$ as a delta function peaking at
\begin{align}
E' \to	\gamma^2\left(\frac{heB_{cr}}{2\pi m c}\right), \nonumber
\end{align}
the observed synchrotron flux can be expressed as \citep{Sahayanathan2018,Shu91} 
\begin{align}\label{eq:synspec}
	F_{\rm syn}(E)\propto E^{-3/2}\rm{exp}\left(-\frac{\xi_0}{\kappa}\,E^{\kappa/2}\right)
\end{align}
Here, $B_{cr}$ is the magnetic field of the CR, $h$ is the Planck constant, $e$ electron charge, $m$ electron rest mass
and $c$ is the velocity of light.
We parameterize the relation between the observed photon energy with the corresponding electron energy through
$\xi_0$ which is given by 
\begin{align}\label{eq:xi0}
	\xi_0=\eta_0\left[\frac{\delta_DheB_{cr}}{2(1+z)\pi m c}\right]^{-\kappa/2}
\end{align}
From equation (\ref{eq:synspec}), 
it is evident the observed synchrotron spectral shape is governed by the parameters $\xi_0$ and $\kappa$ and hence, the 
number of free parameters are equal to the case of log-parabola function. This enables us to perform a detailed 
comparison between these two models based on their X-ray spectral fit results. The peak of the synchrotron spectrum, given by 
equation (\ref{eq:synspec}), in $E^2 F_{syn}(E)$ representation will be 
\begin{align}\label{eq:engtescpeak}
	E_{\rm p,esc} = \left(\frac{1}{\xi_0}\right)^{2/\kappa}
\end{align}
Alternatively, in terms of $E_{\rm p,esc}$ equation (\ref{eq:synspec}) can be expressed as
\begin{align}\label{eq:synspecep}
	F_{\rm syn}(E)\propto E^{-3/2}\rm{exp}\left[-\frac{1}{\kappa}\,\left(\frac{E}{E_{\rm p,esc}}\right)^{\kappa/2}\right]
\end{align}
It can be shown that equation (\ref{eq:synspec}) will be similar to a log-parabola function for $\kappa\ll1$ with the
parameters related as
\begin{align}\label{eq:lpengindex}
	\alpha&\approx\frac{1}{2}[3+\xi_0(1-2.303\,\kappa\,{\rm log}\,E_*)]\\ 
	\label{eq:lpengcurv}
	\beta &\approx 2.303\,\frac{\xi_0\,\kappa}{4}
\end{align}

\begin{figure}
\centering

	\includegraphics[scale=0.8]{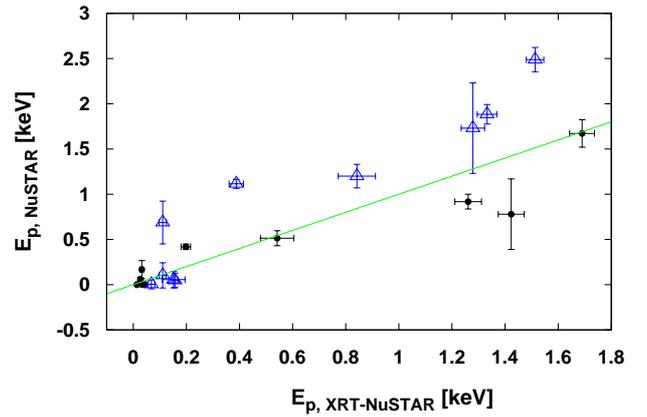}	
	\caption{The plot between SED peaks estimated from the fit parameters corresponding to \emph{Nu}STAR observations alone and the combined \emph{Swift}-XRT and \emph{Nu}STAR observations. Blue triangles corresponds to the log-parabola model and black circles corresponds to the energy-dependent electron diffusion model. The identity line is shown in green.}
	 \label{ep-ep}
\end{figure}

We refit the \emph{Nu}STAR and \emph{Swift}-XRT observations of MKN\,421 by adding equation (\ref{eq:synspec})
as a local model in XSPEC. Similar to the log-parabola model, the 20 hard X-ray observations by \emph{Nu}STAR in 
the energy range 3 -- 79 keV can be well fit by this model. In Table \ref{nustar} (right), we provide the best-fit 
parameters for these observations and the $\chi^2_{\rm red}$ are comparable with the log-parabola model. However, a marginal
improvement in the spectral fit compared to a log-parabola model is obtained for the 10 simultaneous observations by 
\emph{Swift}-XRT and \emph{Nu}STAR. In Table \ref{swnu} (right), we provide the best-fit parameters for the combined 
\emph{Swift}-XRT -- \emph{Nu}STAR observations along with the $\chi^2_{\rm red}$ values and in Figure \ref{engtescplot}, 
we show the broad X-ray spectral fit for high flux state s14 (LHS) and low flux state s8 (RHS). 
Similar to the log-parabola spectral fit, the best-fit parameters for this case differ considerably 
from the fitting of the \emph{Nu}STAR observation alone. Nevertheless, 
the fit is better for the high flux states compared to the log-parabola model. 
In Table \ref{e_p} (right), we provide the SED peaks obtained using the fit parameters corresponding to  \emph{Nu}STAR
alone and \emph{Swift}-XRT -- \emph{Nu}STAR combined spectral fits and their distribution is shown in Figure \ref{ep-ep} (black circles) along with an identity 
line for comparison. 
It can be seen that the distribution is closer to the identity line compared to the case of log-parabola fit, suggesting the
peak estimation is better in the present case.
Hence, it can be contemplated that the functional form given by equation (\ref{eq:synspec}) may represent the broadband
SED of MKN\,421 better than the log-parabola function, though it requires further detailed study to assert this.

\begin{figure}
\centering
	\includegraphics[scale=0.8]{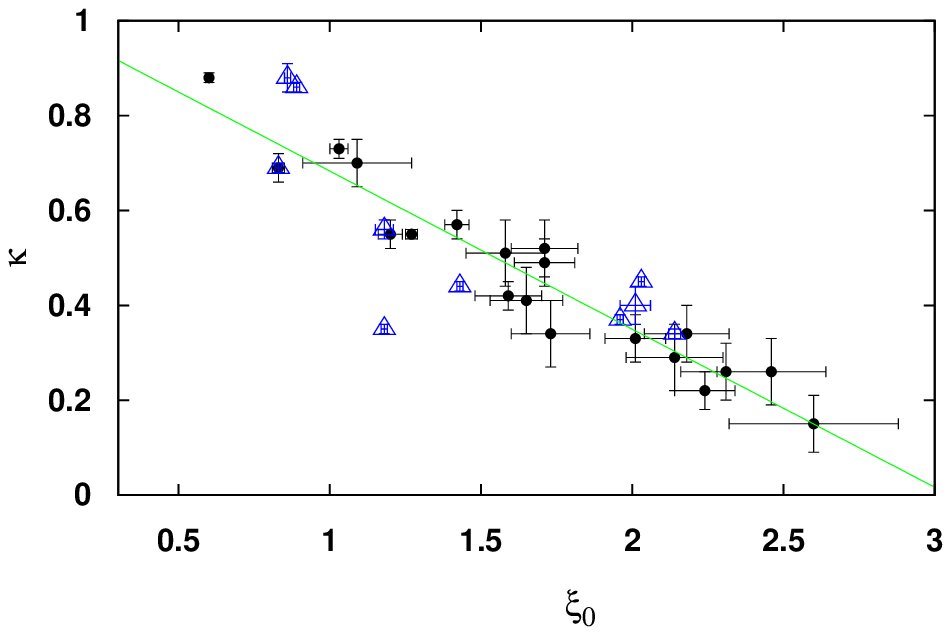}
        \includegraphics[scale=0.8]{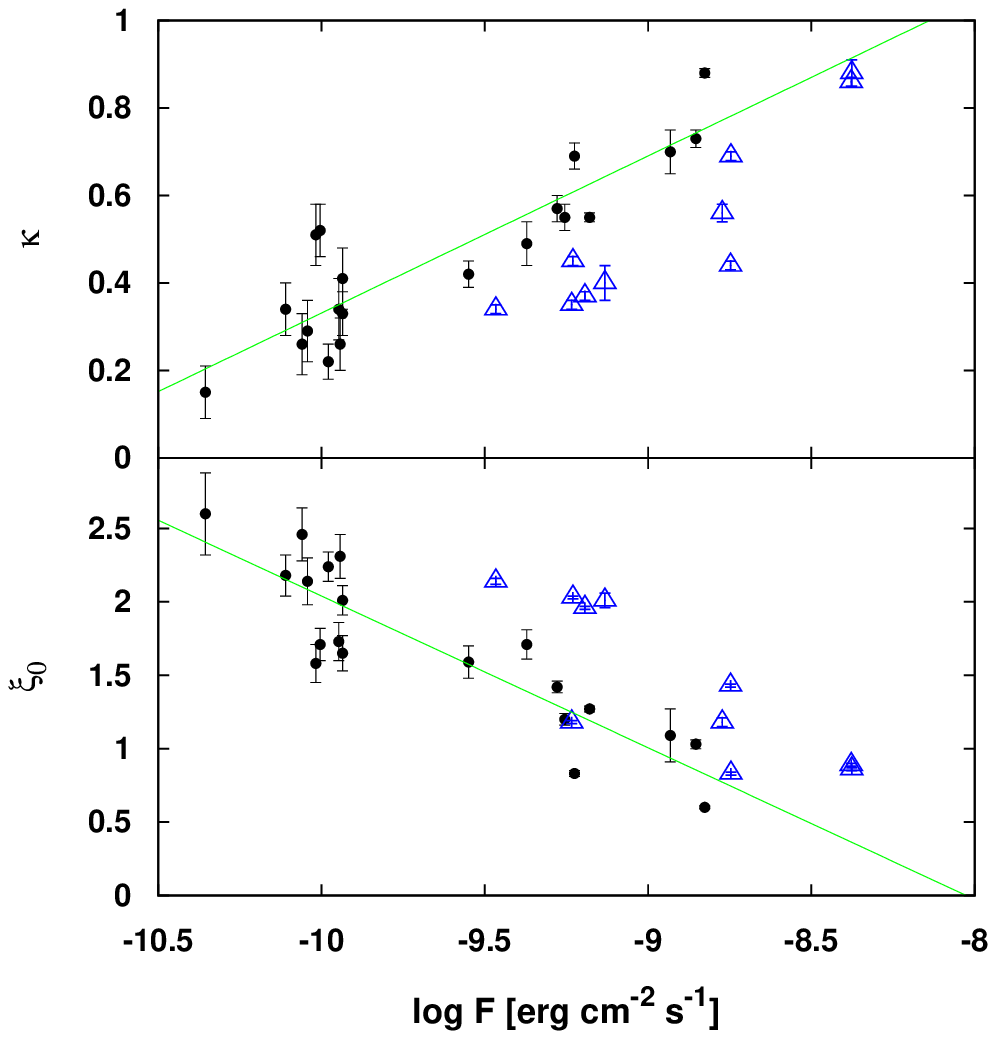}

	\caption{ Scatter plot between the energy-dependent electron diffusion model parameters: $\xi_0$, $\kappa$ and the average flux in the range 3 -- 10 keV and 0.3 -- 10 keV for two different sets of observations. The black circles corresponds to \emph{Nu}STAR observations alone and the blue triangles corresponds to the combined \emph{Swift}-XRT -- \emph{Nu}STAR observations. The solid green lines represent the best-fit for \emph{Nu}STAR observations.}
	\label{eng_corr}
\end{figure}

To study the dependence between the fit parameters and also with the observed properties, we perform a Spearman rank correlation study
among these quantities for the case of log-parabola and the energy-dependent electron diffusion model. In Table \ref{spearman}, we provide the
correlation results where the top rows correspond to log-parabola model and the bottom ones belong to energy-dependent electron diffusion model.
 The scatter plots between the energy-dependent electron diffusion model parameters and the estimated flux are shown in Figure \ref{eng_corr}.
A strong anti-correlation is observed between $\xi_0$ and $F_{3-10\rm\,keV}$ with $r_s=-0.82$ ($p_{\,rs}=9.86\times 10^{-6}$) and a similar
correlation is also obtained for the case of log-parabola parameter $\alpha$ with the correlation coefficient
$r_s=-0.8$ (null probability hypothesis $p_{rs}=3.92\times 10^{-5}$). The latter correlation reflects the ``Harder when brighter" trend
of the source which is consistent with the earlier studies \citep{atreyeeMkn421}. The similarity of the correlation between $\xi_0$ and 
$\alpha$ is expected since they are is closely related through equation (\ref{eq:lpengindex}), particularly for small value of $\kappa$. 
The study between $\kappa$ and $F_{3-10 \rm\,keV}$ showed a strong correlation with $r_s=0.8$ ($p_{\,rs}=2.66\times 10^{-5}$), suggesting 
that the energy dependence of the escape timescale to increase at high flux states. On the contrary, the correlation between  
$\beta$ of the log-parabola model and the $F_{3-10 \rm\,{keV}}$ is moderate 
with $r_s=0.61$ ($p_{rs}=4\times 10^{-3}$). Notably, a strong anti-correlation is seen between the fit parameters $\xi_0$
and $\kappa$ with $r_s=-0.96$ ($p_{\,rs}=1.5\times 10^{-11}$) and this was not the case with the parameters 
$\alpha$ and $\beta$ of the log-parabola model where we obtained a poor correlation with $r_s=-0.39$ ($p_{\,rs}=9.7\times 10^{3}$).
The lack of significant correlation between the log-parabola parameters also suggest that the spectral evolution cannot be attributed 
to the energy dependence of the particle acceleration process proposed in  \cite{Massaro2004a}.
These correlation studies further hint that the functional form given by equation (\ref{eq:synspec}) may probably give a better understanding 
regarding the emission processes in blazars.

\section {Summary and Discussion}
\label{discussion}

In the present work, we perform a detailed X-ray spectral study of the blazar MKN\,421 using \emph{Nu}STAR 
and \emph{Swift}-XRT observations. In addition to modelling the X-ray spectra using a simple log parabola, we use a physical model involving acceleration of 
electrons and subsequent emission through synchrotron radiation. The escape of electrons from the
main acceleration region is assumed to be energy dependent and we found the synchrotron spectrum due to 
this physical model can imitate the curved spectra. We show for 20 epochs of \emph{Nu}STAR observation, a log-parabola 
as well as the energy-dependent electron diffusion models are capable to explain the data very well. However, for 10 epochs 
with simultaneous \emph{Swift}-XRT observations, the energy-dependent electron diffusion model is marginally favoured than the log-parabola.
Moreover, the peak of the SED estimated from the best-fit parameters for the combined  \emph{Swift}-XRT -- \emph{Nu}STAR
and lone \emph{Nu}STAR observations closely satisfy the identity line in case of energy-dependent electron diffusion model compared 
to the log-parabola model. This suggests that the functional form obtained from the energy-dependent electron diffusion model, 
equation (\ref{eq:synspec}), may be capable of representing broadband SED better than the log-parabola model.
Further, the correlation studies indicate that the 
fit parameters of the energy-dependent electron diffusion model are well correlated during various epochs and hence this model is 
capable of deciphering the electron diffusion process in blazar jets.

\begin{figure}
\centering

	\includegraphics[scale=0.8]{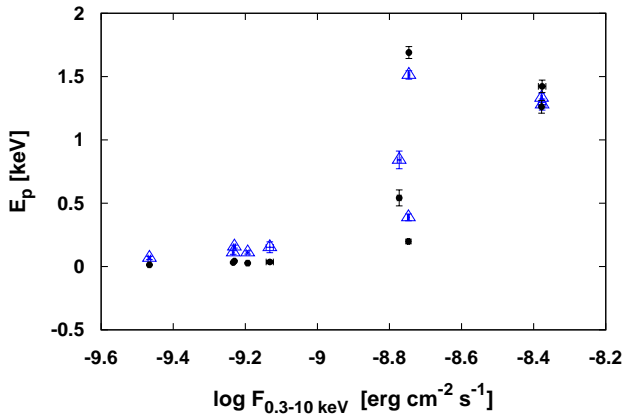}	
	\caption{ The plot between SED peak and 0.3 -- 10 keV flux for the combined \emph{Swift}-XRT and \emph{Nu}STAR observations estimated from log-parabola model (blue triangles) and the energy-dependent electron diffusion model (black circles).}
	 \label{epf}
\end{figure}

The SED peaks estimated using the log-parabola
and the energy-dependent electron diffusion models, $E_{p,\rm\,lp}$ and $E_{p,\rm\,esc}$, through the combined \emph{Swift}-XRT -- \emph{Nu}STAR 
observations are only representative since they sometimes fall at energies lower than the X-ray energy range considered here.
However, they can be used to identify the trend of peak shifts during various flux states.
We computed the Spearman rank correlation between the two estimated spectral peaks and the 0.3 -- 10 keV integrated flux ($F_{0.3-10\rm\,keV}$). The results 
are shown in Table \ref{spearman} and the scatter plot between the peaks and $F_{0.3-10\rm\,keV}$ are shown in Figure \ref{epf}.
We find both $E_{p,\rm\,lp}$ and $E_{p,\rm\,esc}$ correlate well with $F_{0.3-10\rm\,keV}$. The observed positive correlation suggests, the SED peak
shifts towards higher energies during the higher flux states. However, this inference needs to be verified with the information at lower energies.

The analytical solution of the energy-dependent electron diffusion model used in the present work corresponds to the radiatively cooled 
regime of the electron distribution where \textbf{$\gamma\gg (B\,t_e)^{-1}$}.
 This prevents us from extending the current model to UV/optical
energies where the corresponding electron energies may not satisfy this condition. Alternatively, a semi-analytical solution to 
equation (\ref{eq:CR}) can be obtained which will be valid even beyond energies imposed by this condition as well as at the SED peak. However, inclusion
of such a model into XSPEC will be rigorous and will be carried out in a future work. A successful implementation of this will be
effective in constraining $E_{\rm p, esc}$ and can verify the results discussed above. The observational periods of \emph{Nu}STAR and
\emph{Swift}-XRT considered here overlaps partially and are not strictly simultaneous (Table \ref{observation}). This can introduce a substantial 
uncertainty in the estimated parameters owing to the rapid spectral evolution of the source. In this regard, a true simultaneous
observation at energies ranging from UV to hard X-ray bands can be obtained from the multi-wavelength satellite-based 
experiment, \emph{ASTROSAT} which can resolve this discrepancy.

The observed strong anti-correlation between the parameters of the energy-dependent electron diffusion model, $\xi_0$ and $\kappa$, suggests 
these quantities may be dependent on each other. The dependence of $\xi_0$ on $\kappa$ is evident from equation (\ref{eq:xi0}) where the 
latter, besides depending on the physical parameters of the source, is also a function of $\kappa$. On the contrary, if these parameters were found 
to be uncorrelated, then it could have been a sufficient condition to reject the proposed model. Nevertheless, the observed correlation is still not
sufficient enough to validate the assumed model unless the correlation is tested with the choice of parameters that can be shown independently. 
This demands a prior information about the physical parameters of the source (e.g. source magnetic field, jet Lorentz factor, etc.) which cannot be estimated 
from the narrow band of SED considered here. These pieces of information can be obtained through broadband spectral modelling of the source from radio to
 $\gamma$-ray using a numerical solution of the equation (\ref{eq:CR}) and incorporating full synchrotron and inverse Compton emission processes, which will be
 pursued in the future. \\


\textbf{Acknowledgements}: We thank the anonymous referee for insightful comments and constructive suggestions. This research has made use of data, software, and/or web tools obtained from NASAs High Energy Astrophysics  Science Archive Research Center (HEASARC), a service of the Goddard Space Flight Center and the Smithsonian Astrophysical Observatory. This research has made use of the XRT Data Analysis Software (XRTDAS) and the NuSTAR Data Analysis Software (NuSTAR-DAS) jointly developed by the ASI Science Data Center (ASDC, Italy) and the California Institute of Technology (Caltech, USA). PG would like to thank IUCAA and BARC for their hospitality. RG would like to thank the IUCAA associateship programme and CSIR for the financial support.


\begin{table*}
\centering
\footnotesize
\caption{NuSTAR analysis with log-parabola and energy-dependent electron diffusion models}
\label{nustar}
\vspace{0.3cm} 
\begin{tabular}{lcccccccr}
\hline \hline
 Obs. &\multicolumn{3}{c}{Log-parabola} & & \multicolumn{3}{c}{Energy-dependent electron diffusion} &  $F_{3-10\rm\,keV}$ \\
\cline{2-4} 
\cline{6-8} 

& $\alpha$& $\beta$ & $\chi_{\rm red}^{2}({\rm dof})$ &&$\xi_{0}$&$\kappa$&$\chi_{\rm red}^{2}({\rm dof})$ & (10$^{-10}$ erg cm$^{-2}$ s$^{-1}$)
\smallskip \\ 
\cline{1-9}

s1  & 2.96$\pm$0.01 &  0.19$\pm$0.02 &1.09 (801.00)  && 2.24$\pm$0.10 & 0.22$\pm$0.04 &1.09 (802.25)   &  1.05$\pm$0.01 \\
s2  & 2.98$\pm$0.01 &  0.28$\pm$0.02 &1.21 (796.21)  && 2.01$\pm$0.10 & 0.33$\pm$0.05 &1.21 (798.10)   &  1.16$\pm$0.01 \\
s3  & 3.12$\pm$0.01 &  0.31$\pm$0.03 &0.95 (529.99)  && 2.18$\pm$0.14 & 0.34$\pm$0.06 &0.95 (531.59)   &  0.77$\pm$0.01 \\
s4  & 3.07$\pm$0.01 &  0.44$\pm$0.03 &1.01 (575.34)  && 1.71$\pm$0.11 & 0.52$\pm$0.06 &1.01 (575.42)   &  0.99$\pm$0.01 \\
s5  & 2.79$\pm$0.01 &  0.25$\pm$0.03 &1.01 (588.93)  && 1.73$\pm$0.13 & 0.34$\pm$0.07 &1.01 (588.97)   &  1.13$\pm$0.01 \\
s6  & 3.06$\pm$0.02 &  0.14$\pm$0.06 &1.05 (438.26)  && 2.60$\pm$0.28 & 0.15$\pm$0.06 &1.05 (438.21)   &  0.44$\pm$0.01 \\
s7  & 3.07$\pm$0.01 &  0.23$\pm$0.03 &1.04 (578.66)  && 2.31$\pm$0.15 & 0.26$\pm$0.06 &1.04 (579.19)   &  1.14$\pm$0.01 \\
s8  & 3.16$\pm$0.01 &  0.24$\pm$0.04 &0.96 (484.85)  && 2.46$\pm$0.18 & 0.26$\pm$0.07 &0.96 (484.90)   &  0.87$\pm$0.07 \\
s9  & 2.84$\pm$0.01 &  0.31$\pm$0.03 &0.99 (581.57)  && 1.65$\pm$0.12 & 0.41$\pm$0.07 &0.99 (582.02)   &  1.16$\pm$0.01 \\
s10 & 2.80$\pm$0.01 &  0.31$\pm$0.01 &1.04 (933.37)  && 1.59$\pm$0.11 & 0.42$\pm$0.03 &1.05 (935.40)   &  2.82$\pm$0.01 \\
s11 & 3.02$\pm$0.01 &  0.41$\pm$0.02 &1.02 (628.06)  && 1.71$\pm$0.10 & 0.49$\pm$0.05 &1.02 (629.45)   &  4.25$\pm$0.03 \\
s12 & 2.71$\pm$0.01 &  0.37$\pm$0.01 &1.04 (1490.75) && 1.27$\pm$0.02 & 0.55$\pm$0.01 &1.04 (1488.23)  &  6.62$\pm$0.01 \\
s13 & 2.75$\pm$0.02 &  0.49$\pm$0.06 &1.01 (401.57)  && 1.09$\pm$0.18 & 0.70$\pm$0.05 &1.02 (404.32)   & 11.72$\pm$0.12 \\
s14 & 2.71$\pm$0.01 &  0.46$\pm$0.01 &0.91 (921.02)  && 1.03$\pm$0.03 & 0.73$\pm$0.02 &1.22 (1243.05)  & 14.05$\pm$0.15 \\
s15 & 2.89$\pm$0.01 &  0.44$\pm$0.01 &1.00 (910.95)  && 1.42$\pm$0.04 & 0.57$\pm$0.03 &1.01 (925.77)   &  5.27$\pm$0.01 \\
s16 & 2.36$\pm$0.01 &  0.42$\pm$0.01 &1.06 (1516.55) && 0.60$\pm$0.01 & 0.88$\pm$0.01 &1.03 (1481.27)  & 14.90$\pm$0.18 \\
s17 & 2.64$\pm$0.01 &  0.35$\pm$0.01 &0.95 (962.51)  && 1.20$\pm$0.04 & 0.55$\pm$0.03 &0.94 (952.79)   &  5.55$\pm$0.01 \\
s18 & 2.44$\pm$0.01 &  0.36$\pm$0.01 &1.03 (1206.35) && 0.83$\pm$0.02 & 0.69$\pm$0.03 &1.02 (1203.86)  &  5.95$\pm$0.02 \\
s19 & 2.94$\pm$0.01 &  0.40$\pm$0.03 &1.01 (566.86)  && 1.58$\pm$0.13 & 0.51$\pm$0.07 &1.01 (565.86)   &  0.96$\pm$0.01 \\
s20 & 3.00$\pm$0.01 &  0.25$\pm$0.04 &0.90 (467.73)  && 2.14$\pm$0.16 & 0.29$\pm$0.07 &0.90 (468.03)   &  0.90$\pm$0.01 \\ 
                                                                                                                            
\hline \hline
\end{tabular} 
\end{table*}


\begin{table*}
\centering
\footnotesize
\caption{The combined \emph{Swift}-XRT -- \emph{Nu}STAR analysis for 10 epochs using log-parabola and energy-dependent electron diffusion models}
\label{swnu}
\vspace{0.3cm} 
\begin{tabular}{lcccccccccr}
\hline \hline
Obs. &\multicolumn{3}{c}{Log-parabola} && \multicolumn{3}{c}{Energy-dependent electron diffusion} & $F_{0.3-10\rm\,keV}$\\
\cline{2-4}
\cline{6-8}
 
&$\alpha$& $\beta$ & $\chi_{\rm red}^{2}({\rm dof})$ && $\xi_{0}$&$\kappa$&$\chi_{\rm red}^{2}({\rm dof})$ & (10$^{-10}$ erg cm$^{-2}$ s$^{-1}$)
\smallskip \\
\cline{1-9}  \\

s6  & 2.57$\pm$0.01 & 0.24$\pm$0.01  &1.06 (779)  && 2.14$\pm$0.02 & 0.34$\pm$0.01 & 1.20 (779)  &3.41 $\pm$0.03\\
s7  & 2.49$\pm$0.03 & 0.30$\pm$0.02  &1.04 (760)  && 2.01$\pm$0.05 & 0.40$\pm$0.04 & 1.05 (760)  &7.37 $\pm$0.17\\
s8  & 2.52$\pm$0.01 & 0.32$\pm$0.01  &1.48 (921)  && 2.03$\pm$0.01 & 0.45$\pm$0.01 & 1.01 (921)  &5.88 $\pm$0.03\\
s12 & 2.20$\pm$0.01 & 0.24$\pm$0.01  &1.21 (1780) && 1.43$\pm$0.01 & 0.44$\pm$0.01 & 1.28 (1780) &17.88$\pm$0.19\\
s13 & 1.92$\pm$0.03 & 0.34$\pm$0.01  &1.41 (903)  && 0.86$\pm$0.01 & 0.88$\pm$0.03 & 1.22 (903)  &42.02$\pm$1.04\\
s14 & 1.90$\pm$0.01 & 0.39$\pm$0.01  &1.34 (1519) && 0.89$\pm$0.01 & 0.86$\pm$0.01 & 1.30 (1519) &41.86$\pm$0.27\\
s17 & 2.04$\pm$0.01 & 0.29$\pm$0.01  &1.02 (1309) && 1.18$\pm$0.03 & 0.56$\pm$0.02 & 0.98 (1309) &16.85$\pm$0.23\\
s18 & 1.90$\pm$0.02 & 0.25$\pm$0.01  &1.53 (1733) && 0.83$\pm$0.01 & 0.69$\pm$0.01 & 1.28 (1733) &17.93$\pm$0.12\\
s19 & 2.42$\pm$0.01 & 0.22$\pm$0.01  &1.13 (996)  && 1.18$\pm$0.01 & 0.35$\pm$0.01 & 1.09 (996)  &5.83 $\pm$0.31\\
s20 & 2.48$\pm$0.05 & 0.25$\pm$0.01  &1.01 (931)  && 1.96$\pm$0.01 & 0.37$\pm$0.01 & 1.05 (931)  &6.40 $\pm$0.64\\
\hline \hline \\                                      
\end{tabular} 

\end{table*}


\begin{table*}
\centering
\footnotesize
\caption{The estimated SED peaks, $E_{p}$ with the combined XRT-\emph{Nu}STAR (0.3 -- 79 keV)  and \emph{Nu}STAR (3 -- 79 keV) observations using log-parabola and energy-dependent electron diffusion models}
\label{e_p}
\vspace{0.3cm} 
\begin{tabular}{lccccr}
\hline \hline
Obs. &\multicolumn{2}{c}{$E_{p,\rm\,lp}$ (keV)} && \multicolumn{2}{c}{$E_{p,\rm\,esc}$ (keV)} \\
\cline{2-3}
\cline{5-6} 
& XRT-\emph{Nu}STAR& \emph{Nu}STAR &&XRT-\emph{Nu}STAR& \emph{Nu}STAR \\
\cline{1-6}\\

s6  &  0.06$\pm$0.01 &  0.003$\pm$0.05   && 0.01$\pm$0.01  &  1.15e-5$\pm$0.001 \\
s7  &  0.15$\pm$0.04 &  0.05$\pm$0.09   && 0.03$\pm$0.01  &  7.66e-4$\pm$0.01 \\
s8  &  0.16$\pm$0.01 &  0.04$\pm$0.08   && 0.04$\pm$0.01  &  1.05e-3$\pm$0.01 \\
s12 &  0.38$\pm$0.02 &  1.11$\pm$0.05   && 0.19$\pm$0.01  &  0.41  $\pm$0.03 \\
s13 &  1.27$\pm$0.04 &  1.73$\pm$0.50   && 1.42$\pm$0.04  &  0.77  $\pm$0.39 \\
s14 &  1.33$\pm$0.03 &  1.88$\pm$0.10   && 1.26$\pm$0.05  &  0.91  $\pm$0.08 \\
s17 &  0.84$\pm$0.07 &  1.20$\pm$0.13   && 0.54$\pm$0.06  &  0.51  $\pm$0.08 \\
s18 &  1.51$\pm$0.03 &  2.48$\pm$0.13   && 1.69$\pm$0.04  &  1.67  $\pm$0.15 \\
s19 &  0.11$\pm$0.01 &  0.68$\pm$0.23   && 0.03$\pm$0.01  &  0.16  $\pm$0.10 \\
s20 &  0.12$\pm$0.01 &  0.10$\pm$0.13   && 0.02$\pm$0.01  &  0.06  $\pm$0.01 \\

\hline \hline                                       
\end{tabular} \\
\smallskip

\end{table*}

\begin{table*}
\centering
\footnotesize
\caption{Spearman rank correlation results between the model parameters and the observable quantities for log-parabola and energy-dependent electron diffusion model with the \emph{Nu}STAR observations}
\label{spearman} 
\begin{tabular}{lccr}
\hline \hline

 Model&Correlations& r$_{s}$ &p$_{\,rs}$ \\
 
\hline 

&$\beta$ vs $F_{3-10\rm\,keV}$ & 0.61 & 4.0$\times10^{-3}$ \\
Log-parabola&$\alpha$ vs $F_{3-10\rm\,keV}$ &- 0.80 &3.92$\times10^{-5}$\\
&$\beta$ vs  $\alpha$& - 0.39& 9.70$\times10^{-3}$\\
& $E_{p,\rm\,lp}$ vs $F_{0.3-10\rm\,keV}$&0.89 &4.91$\times10^{-4}$\\

\cline{1-4}

&$\kappa$ vs $F_{3-10\rm\,keV}$ &0.80 & 2.66$\times10^{-5}$\\
 Energy-dependent&$\xi_0$ vs $F_{3-10\rm\,keV}$ &- 0.82 &9.86$\times10^{-6}$\\
electron diffusion&$\kappa$ vs $\xi_0$&- 0.96& 1.50$\times10^{-11}$\\
& $E_{p,\rm\,esc}$ vs $F_{0.3-10\rm\,keV}$ &0.89 &5.40$\times10^{-4}$\\

\hline \hline 
\end{tabular} \\
\textbf{Notes}: $E_p$ and $F_{0.3-10\rm\,keV}$ are estimated for the combined \emph{Swift}-XRT and \emph{Nu}STAR \\ observations. Here, r$_s$ is the rank co-efficient and p$_{\,rs}$ denotes the null hypothesis probability
\end{table*}



\bibliographystyle{mnras}
\bibliography{blazar} 

\bsp	
\label{lastpage}
\end{document}